\documentclass[12pt,preprint]{aastex}
\begin{document}

\shorttitle{Reconnection flux emergence}
\shortauthors{Mart\'inez-Sykora \& Hansteen \& Carlsson}

\title{Twisted flux tube emergence from the convection zone to the corona II: Later states}

\author{$^{1}$ Juan Mart\'inez-Sykora}
\email{j.m.sykora@astro.uio.no}
\author{$^{1,2}$ Viggo Hansteen} 
\email{viggo.hansteen@astro.uio.no}
\and 
\author{$^{1,2}$ Mats Carlsson}
\email{m.p.o.carlsson@astro.uio.no}
\affil{$^{1}$Institute of Theoretical Astrophysics, University of Oslo, P.O. Box 1029 Blindern, N-0315 Oslo, Norway}
\affil{$^{2}$Center of Mathematics for Applications, University of Oslo, P.O. Box 1053 Blindern, N-0316 Oslo, Norway}
\newcommand{\myemail}{juanms@astro.uio.no}
\newcommand{\viscous}{\underline{\underline{\tau}}}
\newcommand{\resistive}{\underline{\underline{\eta}}}

\begin{abstract}

Three-dimensional numerical simulations of magnetic flux emergence are 
carried out in a computational domain spanning the upper layers of the convection zone 
to the lower corona. We use the {\em Oslo Staggered Code} (OSC) to solve the full MHD 
equations with non-grey and non-LTE radiative transfer and thermal conduction along
the magnetic field lines. In this paper we concentrate on the later stages 
of the simulations and study the evolution of the structure of the rising flux in the 
upper chromosphere and corona, the interaction between the emerging flux and the weak 
coronal magnetic field initially present, and the associated dynamics. 

The flux tube injected at the bottom boundary rises to the photosphere where it 
largely remains. However, some parts of the flux tube become unstable and expand
in patches into the upper chromosphere. The flux rapidly expands towards the corona, 
pushing the coronal and transition region material aside, lifting and maintaining the 
transition region at heights greater than $5$~Mm above the photosphere for extensive 
periods of time. The pre-existing magnetic field in the corona and transition region 
is perturbed by the incoming flux and reoriented by a series of high Joule heating events.
Low density structures form in the corona while at later times
a high density filamentary structure appears in the lower part of the expanding flux.
The dynamics of these and other structures is discussed.
While Joule heating due to the expanding flux is 
episodic, it increases in relative strength as fresh magnetic field rises and becomes 
energetically important in the upper chromosphere and corona at later times. 
Chromospheric, transition region and coronal lines are computed and their response to 
the perturbation caused by the expanding emerging flux is discussed.

\end{abstract}

\keywords{Magnetohydrodynamics MHD ---Methods: numerical --- Radiative transfer --- Sun: atmosphere --- Sun: magnetic field --- Sun:photosphere --- Sun: chromosphere --- Sun: granulation --- Sun: corona}

\section{Introduction}
\label{sec:introduction}

Current understanding of the outer solar atmosphere hinges on the active role of the 
solar magnetic field in controlling the dynamics and the energetics of the chromosphere 
and corona. Even outside of active regions the quiet and semi-quiet photosphere is 
threaded by magnetic fields that can appear as bright points, darker pores or
micro-pores \citep{Berger:2004ij,Rouppe-van-der-Voort:2005th}. Magnetic structures are 
subject to photospheric motions which do work on the magnetic field, resulting in a 
Poynting flux that propagates into the chromosphere and corona, presumably dissipating 
and heating the tenuous outer atmosphere. In addition to photospheric stressing, the 
extant magnetic field is also subject to fresh new field emerging through the photosphere, 
having been formed below, somewhere in the solar interior.  Flux emergence is a 
relatively slow process, in the Sun taking of order hours to penetrate the photosphere
and enter the upper atmosphere (as discussed by \citet{paper1}, hereafter referred to 
as Paper~I). Flux emergence is also a large scale process; active regions cover an area 
of some 100$\times$100~arcsec$^2$, while even smaller scale quiet Sun emergence covers 
several granules, say 10$\times$10~arcsec$^2$. The long timescales and the large physical 
dimensions of flux emergence mean that numerical simulations of this process are 
challenging, especially if one is interested in studying emergence all the way into the 
corona, where the emerging field rapidly balloons horizontally in size to fill a 
considerable volume \citep{archontis2004}. 

The magnetic field held in the overshoot region of the convection zone in the solar interior 
undulates and is subject to instabilities that lead to the formation of magnetic flux tubes,
which rise up to the solar surface on the time scale of weeks. However, the rise speed of 
a flux tube is strongly dependent on its twist and magnetic field strength 
\citep{Kosovichev:2008mi,Jouve:2007oz}. A certain minimum twist is required to suppress 
the fragmentation of a flux tube rising buoyantly through the convection zone
\citep{moreno1996,cheung2007,paper1}. In the photosphere the plasma is convectively 
stable and the rising flux tube stalls, until a sufficient gradient in the field has built 
up to trigger a buoyancy instability that allows (portions of) the flux tube to expand into 
the upper atmosphere \citep{archontis2004}. 

The effects of flux emergence are followed through the chromosphere and into the corona in Hinode 
observations at small scales in the quiet Sun \citep{Hansteen:2007gf}.
Hinode observations also show upflows that resemble buoyant plumes in the vicinity of 
quiescent prominences \citep{Berger:2008qf}. These recent observations show the importance 
of the connectivity between all the layers above the upper convection zone. The same can 
be said for recent efforts at simulating the solar atmosphere  
\citep{Gudiksen:2005lr,abbett2007,paper1}. Thus, it is clear that in studying the physics 
of the chromosphere and corona and, in particular, the evolution of the emerging magnetic 
field as it rises into these regions, it is vital to include all layers from the upper 
convection zone to the corona.

In Paper~I we describe a numerical simulation of the initial stages of the emergence of a 
magnetic flux tube into a model solar atmosphere. We found various atmospheric responses 
to the perturbations caused by the flux tube. In the photosphere the granular size
increases as the flux tube approaches from below, as previously reported in the 
literature. In the convective overshoot region, 
some 200~km above the photosphere, adiabatic expansion of the emerging plasma produces 
cooling, and dark regions with the structure of the perturbed granulation cells. 
Collapsing granulation cells are frequently found close to the boundaries 
of the rising flux tube. Once emerging flux has crossed the photosphere, bright points 
related with concentrated magnetic field, high vorticity, high vertical velocities and 
heating by compressed material are found at heights up to 500~km above the photosphere. 
At greater heights, in the magnetized chromosphere, the emerging flux produces a large, 
cool, dim, magnetized bubble that tends to expel the usually prevalent chromospheric 
oscillations \citep{Hansteen:2007gf}. The rising flux also dramatically increases the 
chromospheric scale height, pushing the transition region and corona aside such that 
the chromosphere extends up to 6~Mm above the photosphere in the region of flux 
tube emergence. The emergence of the magnetic flux tube through the photosphere and 
the subsequent expansion through the chromosphere up towards the lower corona is a 
relatively slow process, taking of order 1~hour. 

In this paper our goal is to study the subsequent interaction of an 
emerging, initially horizontal flux tube with the outer portions of the solar
atmosphere in a realistic 3D MHD model. In short, the models discussed here
(and in Paper~I) include: radiative losses from the photosphere and lower chromosphere 
computed in a non-grey manner with scattering in the chromosphere, parametrized 
radiative losses in the upper chromosphere and corona and thermal conduction in the 
hot coronal plasma along magnetic field lines. Horizontal flux tubes with varying 
strength and degrees of twist are injected through the bottom boundary into an atmosphere 
that contains a pre-existing, but weak, magnetic field. 

The MHD equations including radiative transfer, thermal conduction, viscosity and 
resistivity 
as solved by the {\em Oslo Staggered Code} are presented in section~\ref{sec:equations}.  
Section~\ref{sec:condition} gives a short description of the initial and boundary conditions 
for the different simulations. The results of our simulations are 
described in section~\ref{sec:results}, where we discuss the magnetic field evolution, 
the Joule heating that occurs in the atmosphere, the structure and dynamics and the total 
emission of chromospheric and coronal lines. A short discussion and our conclusions are
presented in section~\ref{sec:conclusions}.

\section{Equations and Numerical Method}
\label{sec:equations}

In order to model rising magnetic flux tubes through the upper convection layer and their 
emergence into the photosphere, the chromosphere and corona we solve the equations 
of MHD using the {\it Oslo Stagger Code} (OSC):

\begin{equation}
\frac{\partial \rho}{\partial t} + \nabla ({\rho \bf u})=0, \label{eq:cont}
\end{equation}
\begin{equation}
\frac{\partial {\bf u}}{\partial t}  + ({\bf u} \cdot \nabla) {\bf u} + \frac{1}{\rho} \nabla (P+\viscous) =
	 \frac{{\bf J} \times {\bf B}}{\rho}+{\bf g},\label{eq:moment} 
\end{equation}
\begin{equation}
\frac{\partial (e)}{\partial t}  + \nabla \cdot (e{\bf u})  +  P \nabla \cdot {\bf u}=
	 \nabla \cdot {\bf F}_r+\nabla \cdot {\bf F}_c+Q_{Joule}+Q_{visc},\label{eq:ener} 
\end{equation}
\begin{equation}
\frac{\partial{\bf B}}{\partial t}=\nabla \times ({\bf u} \times {\bf B})-\nabla\times (\resistive {\bf J}), \label{eq:induction}
\end{equation}

\noindent where $\rho$ represents the mass density, ${\bf u}$ the fluid velocity, $P$ the
gas pressure, ${\bf J}$ the current density, ${\bf B}$ the magnetic field, ${\bf g}$ 
the gravitational acceleration, and $e$ the internal energy. The viscous stress tensor is 
written as $\viscous$ and the resistivity as $\resistive$ (more details  in section~\ref{sec:vis}). ${\bf F}_r$ represents the radiative flux, 
${\bf F}_c$ the conductive flux, while $Q_{Joule}$ and $Q_{visc}$ are the Joule 
heating and the viscous heating, respectively. 

The method of solution is presented in some detail in Paper~I, here we will present the method
of treating the viscosity and resistivity since Joule heating is a central theme in
this paper. 

\subsection{Viscosity and resistivity}
\label{sec:vis}

The OSC utilizes a high order artificial diffusion to suppress the numerical 
noise inherent in a finite difference scheme. Diffusion is a second order
operator; however, it has been shown that with a higher order operator one 
can suppress the unwanted perturbations with wavelengths shorter than 
$2\pi\Delta x$ (where $\Delta x$ is the grid spacing) while well resolved 
physical structures are damped less than with second
order diffusion. We therefore follow Nordlund, \AA~ \& Galsgaard, K 1995 
(see {\tt http://www.astro.ku.dk/$\sim$kg}). They define a
higher order ``hyper-diffusion'' operator in the following manner:
\begin{equation}
d_i(f)=q_i(\partial_{i}(f))\partial_{i}(f)
\end{equation}
\noindent
where the so called ``quenching'' operator ($q$) is given by:
\begin{equation}
q_i(f)=\frac{\max_{i\pm1}|\Delta_2 f|}{\max_{i\pm1}|\Delta_1 f|}.
\end{equation}
\noindent
The {\it max} is the maximum over three grid points in the derivative direction $i$, 
and $\Delta_1$ and $\Delta_2$ represent first and second order differences, respectively.
This operator is of order unity for high wavenumber perturbations and 
diminishes as $k^{-2}$ for smaller wavenumbers. 

Using these operators the stress tensor is defined as: 
\begin{equation}
\tau_{ij} = \rho \frac{1}{2}\left[(\nu^{(1)}_i+\nu^{(1)}_j)+
	(\nu^{(2)}_i+\nu^{(2)}_j)\right]\left(d_{j}(u_i)+d_{i}(u_j)\right)
\end{equation}

\noindent In order to deal with numerical instabilities that arise as a result
of transport errors, shocks and other high gradient phenomena on a grid of 
finite numerical resolution we use:

\begin{eqnarray}
\nu^{(1)}_j=\Delta x_j(v_1c_f+v_2|u_j|), \\
\nu^{(2)}_j= \Delta x_j^2(v_3|\nabla\cdot{\bf u}|_{-}),
\end{eqnarray}

\noindent where $c_f$ is the fast mode speed, $v_1$, $v_2$ and $v_3$ are 
dimensionless numbers of order unity, and $|...|_{-}$ denotes the 
absolute value of the negative part of the operator which is zero otherwise,
such that $\nu^{(2)}$ is non-zero only during compression.

In a similar manner, an artificial diffusivity is included to deal with numerical 
instabilities associated with the magnetic field. As the stresses 
in the coronal field grow, so does the energy density of the field. This energy 
must eventually be dissipated at a rate commensurate with 
the rate at which energy is pumped in. In the Sun the magnetic diffusivity 
$\eta$ is very small and gradients must become very large before dissipation 
occurs. In the models presented here we operate with an $\eta$ many orders 
of magnitude larger than on the Sun and dissipation starts at much smaller 
magnetic field gradients. The dissipated energy is:

\begin{equation}
Q_{Joule}={\bf E}\cdot{\bf J},
\label{eq:joule}
\end{equation}

\noindent where the resistive part of the electric field is given by:

\begin{equation}
E_x^{\eta}=\left\{{1\over 2}\left(\eta^{(1)}_yq_y(J_x)+\eta^{(1)}_zq_z(J_x)\right)
    +{1\over 2}(\eta^{(2)}_y+\eta^{(2)}_z)\right\}J_x,
\end{equation}

\noindent and similar for $E_y^{\eta}$ and $E_z^{\eta}$. The diffusivities are 
given by:

\begin{eqnarray}
\eta^{(1)}_j={\Delta x_j\over{\rm Pr_M}}(v_1c_f+v_2|u_j|), \nonumber \\
\eta^{(2)}_j={\Delta x_j^2\over{\rm Pr_M}}(v_3|\nabla_\perp\cdot{\bf u}|_{-})
\label{eq:eta}
\end{eqnarray}

\noindent where ${\rm Pr}_M$ is the magnetic Prandtl number, that is the ratio of viscous to magnetic diffusion and $\nabla_\perp{\bf u}$ is the divergence 
of the velocity field perpendicular to the magnetic field. Note that the 
latter term, as opposed to the hydrodynamic shock viscosity $\nu^{(2)}$, 
does not vanish everywhere in incompressible flows. It is instead large enough 
to halt the collapse of a magnetic flux tube in a region of perpendicular 
convergence near the limit of the numerical resolution.
 
Finally, the diffusion term of the magnetic field is:

 \begin{eqnarray}
 \resistive{\bf J} = \frac{1}{2}\sum_{i,j}\left[(\eta^{(d)}_i+\eta^{(d)}_j)\right]
    (\frac{\partial B_i}{\partial x_j}-\frac{\partial B_j}{\partial x_i}),
 \end{eqnarray}

 \noindent where:

 \begin{equation}
 \eta^d_i=\eta^{(1)}_i+\eta^{(2)}_i
 \end{equation}

\section{Initial and boundary conditions}
\label{sec:condition}

The models described here have a grid of $256\times 128\times 160$
points spanning a solar volume $16\times 8\times 16$~Mm$^3$. The bottom boundary is
situated $1.4$~Mm below the photosphere (which we define to be at $z=0$). We have chosen a uniform grid 
spacing of $65$~km in the $x$ and $y$ directions. The non-uniform grid adopted 
in the $z$ direction ensures that the vertical resolution is good enough to resolve 
the photosphere and the transition region with a grid spacing of $32.5$~km, while becoming 
larger at coronal heights.

Note that the atmosphere is very dynamic; the chromosphere, transition region and 
corona do not have borders at constant heights and the corrugated surfaces evolve in time; {\it e.g.,} see
figure~1 of Paper~I. When discussing the results we therefore give the explicit heights
of horizontal cuts and only sometimes an indication of the physical regime that
dominates a given height range.

We have seeded the initial model with two different magnetic field configurations. In one
sufficient stresses can be built up to maintain coronal temperatures in the upper 
part of the computational domain, as previously shown to be feasible by 
\citet{Gudiksen+Nordlund2004}. In the other models, the field is weaker and the corona slowly
cools as the energy derived from foot point braiding is insufficient to maintain the initial coronal temperature. 
In these latter simulations (simulation A1 and A2) the average unsigned field strength in the photosphere
is only $16$~Gauss and we find fairly weak coronal heating and temperatures of
order $5\,10^5$~K and falling, while in the other (simulation B1) the average unsigned field strength
is $155$~Gauss and the coronal temperature is of order $1$~MK or higher. The initial field in both
cases was obtained by semi-randomly spreading some $20-30$ positive and negative 
patches of vertical field at the bottom boundary,  and
then calculating the potential field that arises from this distribution in the remainder of 
the domain. The field is advected by convective flows and photospheric motions and stresses 
sufficient to maintain a minimal corona which is built up by photospheric motions after 
roughly $20$~minutes solar time. 

\begin{figure*}
  \includegraphics[width=0.98\textwidth]{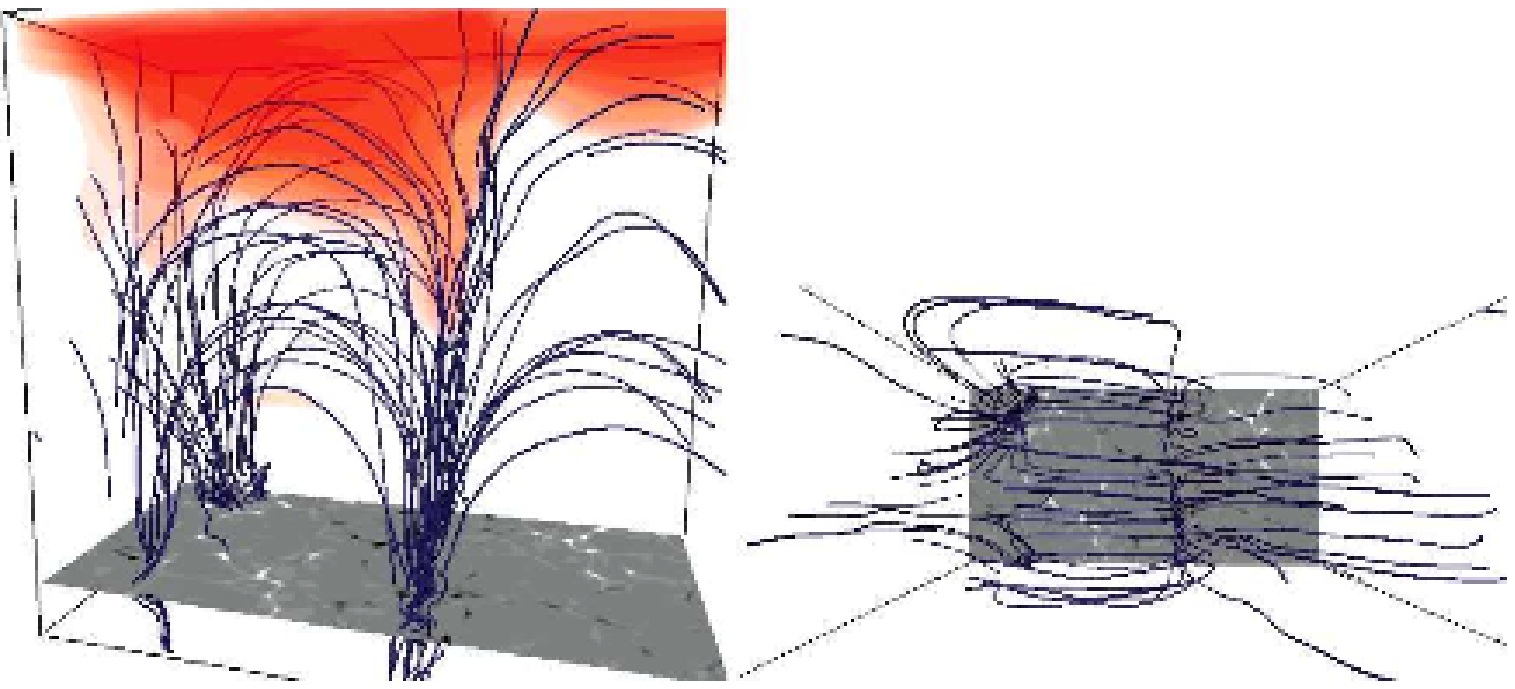}
  \caption{\label{fig:3dri} Computational box from the convection zone to the corona shown at 
  time $500$~s as seen from the side (left panel) and from above (right panel). Lowest density is shown in red; compare that region with later stages (see figure~\ref{fig:3dr}). Magnetic field lines (blue) are mostly oriented along the $x$ axis. The vertical magnetic field in the photosphere is shown with a grey-scale layer.}
\end{figure*}

During this time we also find that the field topology is organized by the flow. After 
relaxation of the original field, the coronal loops in both sets of models are generally aligned 
along the $x$-axis, stretching from magnetic field concentrations centered roughly at $x=7$~Mm and 
$x=13$~Mm. The initial field geometry is shown in figure~\ref{fig:3dri}, where a subset 
of the field lines reaching the corona are shown from the side (in the left panel) and from 
above (right panel). Also shown in this figure is the distribution of the vertical magnetic field
$B_z$ that pierces the photosphere (greyscale) and the regions of lowest mass density (in red), 
which in the initial model are concentrated near the top of the model, as expected. We shall 
later see that regions of low density can occur at lower heights after flux emergence has 
penetrated the corona.

\subsection{Initial flux tube}
\label{sec:initial-model}

We introduce a magnetic flux tube into the lower boundary of the model 
as described in detail in section~3.2 of Paper~I.
The magnetic flux tube structure is horizontally axisymmetric
and the longitudinal field has a gaussian profile given by:

\begin{eqnarray}
{\bf B}_{l}&=& B_o\, \exp \left(-\frac{r^2}{R^2}\right)\, {\bf e}_{l}, \label{eq:blong}\\
{\bf B}_{t}&=&B_{l}\,r\,q\,{\bf e}_{\phi} ,\label{eq:btrans}
\end{eqnarray}

\noindent where  $r=\sqrt{(x-x_o)^2+(z-z_o)^2}$ is the radial distance to the center of the 
tube and $R$ is the radius of the tube. ${\bf B}_{l}$ and ${\bf B}_{t}$ are the longitudinal and 
transverse magnetic fields in cylindrical coordinates, respectively. The 
parameter $q$ is used by \citet{linton1996} and \citet{fan1998} to define the twist of the 
magnetic field. Following \citet{mark2006}, we define a dimensionless twist parameter, $\lambda$, as:

\begin{equation}
\lambda\equiv q\,R. \label{lambda}
\end{equation}

As the flux tube enters the computational box, the height of the center of the tube 
($z_o$) changes in time. The speed of the flux tube 
$({dz_o/dt})$ is set to the average of the velocity of the plasma inflow at the boundary 
in the region where the magnetic flux tube is located at each time step. 

We have made three simulations (A1, A2, and B1): 
The emerging flux tube in runs A1 and A2 has the same strength, 
$B_0=4500$~Gauss at the bottom boundary, but the twist differs, $\lambda=0.3$ in simulation 
A1 and $\lambda=0.6$ 
in run A2. The latter simulation, B1, has a stronger pre-existing ambient field than the two 
others, while the injected field is weaker, $B_0=1100$~Gauss at the bottom boundary. The tube
has no twist but a larger radius than simulations A1 and A2, and could indeed be described 
better as a flux slab than a flux tube.
The emerging flux tubes injected at the bottom boundary are oriented in the $y$-direction 
and are centered roughly at 8~Mm in all three cases. A summary of the runs completed is shown 
in table \ref{tab:runs}. The magnetic energy 
of the tube is only a few times higher than the equipartition energy 
(defined as the average of the kinetic energy of the convective flows).
The flux tube is weak enough to be perturbed and fragmented 
by the convective motions in all three models (in order to have a coherent
tube the magnetic field energy would have to be more than 10 times more than
the equipartion energy \citep[{\it e.g.,}][]{Fan:2003ad,cheung2007}. We don't
discuss that case here).

\section{Results}
\label{sec:results}
 
In this paper we specifically concentrate on the later stages of the simulations where we
study the evolution of the fields as it rises into the upper chromosphere 
and corona, the interaction between 
the field and the weak coronal magnetic field initially present, 
and the associated dynamics. In the following sections all the figures shown 
correspond to simulation A2 unless otherwise 
noted. This both because the A2 simulation has been run for the longest period of solar time
($> 5000$~s) and, in addition, as it has the largest
amount of magnetic flux emerging and passing into the outer layers. 

The early stages of the flux emergence were covered in Paper~I; for a summary see 
the introduction (section~\ref{sec:introduction}).
Let us continue to follow the expansion of the field into the 
upper chromosphere and lower corona. As the magnetic flux moves into the upper 
chromosphere, it expands into the corona, pushing the coronal and transition region material
aside, lifting and maintaining the transition region at heights greater than $5$~Mm above
the photosphere for extensive periods of time. The cool magnetized bubbles reported in the 
chromosphere in Paper~I gradually become elongated in the direction of the tube as they become
weaker. They continue to thin until they disappear, and  shock waves once again dominate the
chromospheric dynamics. The pre-existing magnetic field in the 
corona and transition region is perturbed by the incoming flux and reoriented by a series 
of interactions between the two systems. A high density filamentary structure appears in the lower part of 
the flux tube and in regions below. While Joule heating is 
episodic, it increases in relative strength as fresh magnetic field rises and becomes 
energetically important in the upper chromosphere and corona at later times. All these 
processes will be described in the following sections.

\subsection{The magnetic field}
\label{sec:structure}

The only parts of the flux tube that can expand into the layers above the photosphere are those 
where the gradient of the magnetic field strength is larger than the superadiabatic excess 
in the photosphere. This relation between the buoyancy instability and the superadiabatic
excess have been studied by several authors \citep{Acheson:1979lr,magara2001,archontis2004}. In \citet{cheung2007} and in Paper~I, it is pointed out that a greater amount of 
twist in the initial flux tube allows a larger fraction of magnetic flux to pass
into the regions above the photosphere. 
The remaining parts of the magnetic field stay in the photosphere and move with the fluid
into intergranular lanes where, eventually, the field is either pulled down into the 
convection zone or establishes enough of a gradient to be able to rise into the outer 
atmosphere. The expansion into the chromosphere does not happen uniformly, but rather 
in patches of several Mm$^2$ area. These regions of strong field remain in the same 
location in the chromosphere for several minutes and are related to 
the cold regions in the chromosphere described in Paper~I. 

\begin{figure*}
  \includegraphics[width=0.98\textwidth]{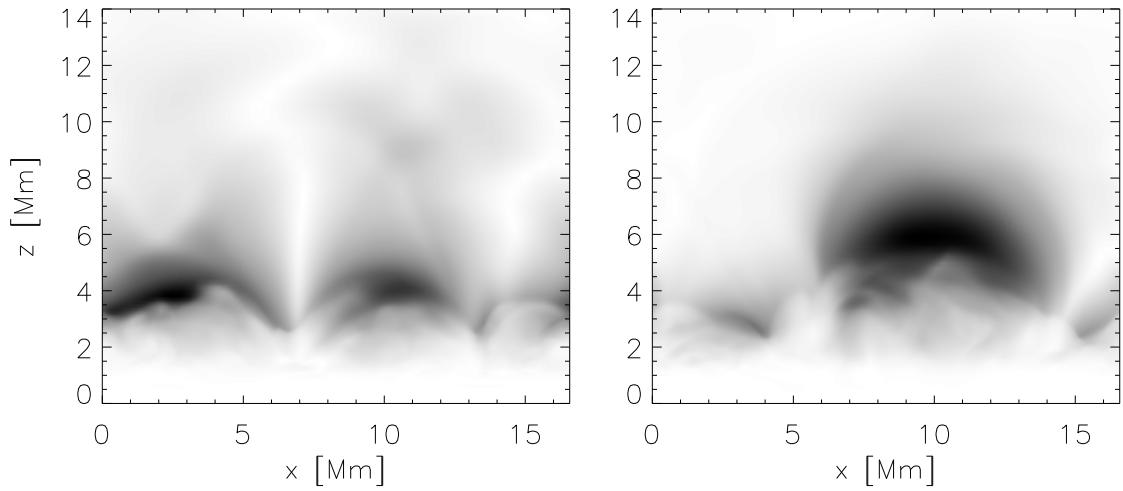}
 \caption{\label{fig:bycor} Vertical cut at $y=650$~km of the horizontal magnetic 
field strength normalized with the square root of the mean over the $y$ axis of the density. Two snapshots are shown: before the tube crosses the photosphere at time $600$~s (left-panel) and  after the tube crosses the photosphere
at $4200$~s (right panel). The panels are individually scaled: at time $4200$~s 
the maximum strength of the horizontal field is 3 times greater than at time $600$~s.}
\end{figure*}

\begin{figure*}
  \includegraphics[width=0.98\textwidth]{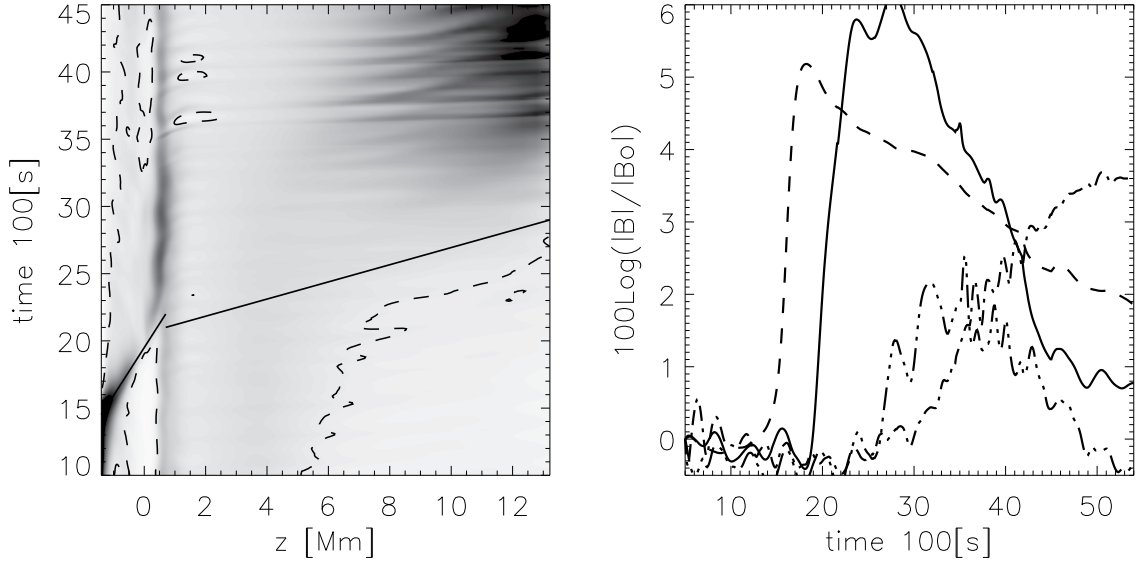}
 \caption{\label{fig:bratio} Ratio between the absolute horizontal and vertical magnetic 
field strength as averaged over the {\em x} and {\em y} axes as a function of time and height (left panel, see equation~{\ref{eq:bhzrat}}).
The minimum value of the ratio is $7\,10^{-5}$ (white), while the maximum is $22.5$ (black). The regions where the ratio
is greater than $1$ are delineated by the dashed lines.
The average emergence velocity of the tube in the convection zone and the average expansion 
velocity of the tube above the photosphere are shown with straight solid lines. The right panel 
shows the average of the absolute magnetic field strength
over specific height intervals and normalized with the same average taken over the first $500$~s
of the simulation, see equation~\ref{eq:bborat}, in the upper convection zone
($z=[-854,-150]$~km, dashed line), the upper photosphere ($z=[106,490]$~km, solid line), 
the chromosphere ($z=[650,1130]$~km, dot-dot-dot-dashed line) and the corona ($z=[4108,12300]$~km, dot-dashed line).}
\end{figure*}

The long term evolution of the coronal magnetic field is illustrated in figure~\ref{fig:bycor} by 
showing the horizontal field at two times: before (left panel) 
and well after (right panel) the chromospherically 
buoyant regions of the tube have reached the corona (see also figure~\ref{fig:bratio}). 
In runs with no flux emergence the lower coronal field structure shown in the left panel of 
figure~\ref{fig:bycor} has a lifetime of several hours, since the field that reaches to
coronal heights is rooted well below the photosphere, where the timescales are correspondingly 
long. {\it I.e.} in the absence of flux emergence we would expect this general coronal field 
topology to be unvarying on timescales much longer than the simulations presented here. 
In the present case we do have flux emergence and the emerging flux tube's axis reaches the
photosphere at roughly 2000~s and remains in the vicinity of the 
photosphere/lower chromosphere until the end of the simulations \citep[see also][]{fan2001,magara2003,archontis2004,manchester2004,Murray:2006ly,Galsgaard:2007mz}. 
However, a portion of the tube expands upward and enough flux is transported to the 
upper-chromosphere to change the coronal field topology.
This expansion happens at a fairly slow velocity, that decreases in time, 
but by $t=4200$~s figure~\ref{fig:bycor} shows that the emerged flux completely dominates the upper 
chromosphere and lower corona.

Although it is almost impossible to follow the axis or the apex of the flux tube in these 
simulations because of the fragmented nature of the tube and its interaction with the 
ambient magnetic field. However, the rise of the tube and its subsequent expansion into the 
corona is nicely illustrated by following the evolution of the ratio between the horizontal 
and vertical magnetic field strength with height and time:

\begin{eqnarray}
\frac{\langle |B_h|\rangle_{xy}}{\langle |B_z|\rangle_{xy}} \label{eq:bhzrat} 
\end{eqnarray}

\noindent where $\langle \cdots\rangle_{xy}$ is the mean over $x$ and $y$ (see left panel of figure~\ref{fig:bratio}). 
We see that in the convection zone the flux tube rises at an average velocity of $3$~km s$^{-1}$. 
This velocity is similar to the mean upflow velocity of the convective motions but should
also be compared with the buoyant velocity of the flux tube.
The buoyant velocity 
is chosen as the terminal velocity between the buoyant force and the resistivity force:

\begin{eqnarray}
u_t=\sqrt{\frac{\pi R g }{C_D}\frac{\Delta\rho}{\rho}},
 \end{eqnarray}

\noindent where $C_D$ is the drag 
coefficient and $\Delta \rho$ is the difference of the density inside the tube and its surroundings. 
We have chosen $C_D= 2$ which seems suitable for our models. Assuming
pressure equilibrium between the inside and outside of the tube we obtain

\begin{eqnarray}
u_t \approx \sqrt{\frac{\pi R g }{C_D}\frac{1}{\gamma\beta}}.
 \end{eqnarray}

\noindent For simulations A1 and A2 this expression gives a buoyant velocity $u_t\approx 3$~km/s.

The expansion of the flux tube into the corona can also be followed in figure~\ref{fig:bratio}.
The mean velocity of the expansion is roughly $16$~km s$^{-1}$, similar to the velocity of $17$~km s$^{-1}$
reported by \citet{archontis2004}. We also see an oscillatory wavelike pattern. Phases of large relative 
horizontal field seem to propagate into the corona like the rings that propagate when a stone is dropped into 
a pond. The horizontal field phase velocity is similar to the slow mode speed 
(around $30$~km s$^{-1}$ in the upper chromosphere) and at greater heights the phase speed
can be seen to split to follow both slow mode and Alfv{\'e}n speeds. The period of the 
oscillations is roughly $5$ minutes. Note that the field in the corona becomes steadily 
more horizontal as time progresses.

Looking in detail at the left panel of figure~\ref{fig:bratio} we see that the ratio between 
horizontal and vertical field decreases 
as the flux tube gets closer to the photosphere ($z\approx 0$~km) from below. However, once the 
tube reaches the photosphere there is an increase of the relative horizontal field
as the tube field ``piles up''. Some $700$~s after the tube reaches the photosphere 
the ratio between horizontal and vertical field begins decreasing slowly due to
flux expansion into the corona and some submergence of field back 
into the convection zone. 

The right panel of figure~\ref{fig:bratio} shows the evolution of flux emergence as function of
height and time. To show the global behavior we take the average of the absolute magnetic field strength
over specific height intervals and normalize with the same average taken over the first $500$~s
of the simulation:

\begin{eqnarray}
&&\frac{\langle |B|\rangle_{xy}(z,t)}{\langle |B|\rangle_{xyt[0:500]}(z)} \label{eq:bborat}
\end{eqnarray}

\noindent where $\langle \cdots\rangle_{xyt[0:500]}$ is the mean over $x$, $y$ and $t=[0:500]$~s. 
This procedure removes the pre-existing field and highlights the emerging flux. We show the
evolution of this quantity as a function of time for four different height intervals: the
upper convection zone (dashed line), the upper photosphere (solid line), 
the chromosphere (dot-dot-dot-dashed line) and the corona (dot-dashed line). Note the rapid
increase and much more gradual decrease in the convection zone and photosphere and the 
more gradual emergence into the chromosphere and corona.

\begin{figure*}
  \includegraphics[width=0.98\textwidth]{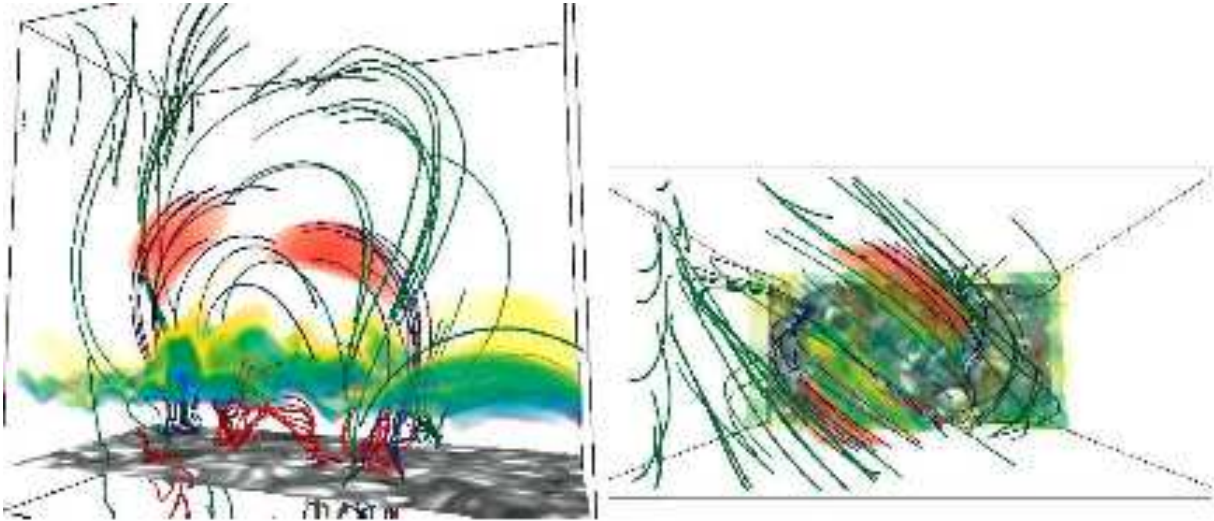}
  \caption{\label{fig:3dr} Computational box from the convection zone to the corona at 
  time  $4500$~s as seen 
  from the side (left panel) and from above (right panel). Density is shown in color:
  blue-green-yellow-red in order of decreasing density 
  (compare with figure~\ref{fig:3dri}). Magnetic field lines are drawn from regions
  of low density (blue lines), high density regions (red lines) and
  where the magnetic 
  field was open $400$~s before (green). The temperature in the photosphere is shown with 
  a grey-scale layer.}
\end{figure*}

It is interesting to follow the evolution of the chromospheric and coronal 
field as the flux tube expands into the outer atmosphere.  In 
figure~\ref{fig:3dr} three sets of field lines are shown at $t=4500$~s. 
By this time the upper part of the expanding flux tube has risen into the corona 
and the associated field lines are drawn in blue. As these field lines 
rise, material drains along the field and a low density region is formed 
(shaded in red in figure~\ref{fig:3dr}). In response, the previously open
field lines above the low density region (drawn in green) close and are pulled 
downwards. Not all of the expanding field makes it up into the 
corona; below the coronal portions of the rising flux we find field lines, 
(drawn in red) that remain in the chromosphere and are subject to the
forced dynamics of that region. The highest (green) magnetic field lines are nearly 
horizontal near their apex, while the lower lying (blue) field lines have shapes 
that are much rounder. 

Note that the three sets of field lines shown in figure~\ref{fig:3dr} connect to different locations in the photosphere. 
The photospheric foot points of the coronal and chromospheric field lines are well concentrated
but to different locations. By analyzing the velocity pattern at the bottom boundary, 
1.4~Mm below the photosphere, we find that we can define a ``mesogranular'' scale that stretches 
over a scale equivalent to some $3-5$ granules horizontally. 
Field lines that are brought up by convection and/or 
in the rising flux tube first appear in the center of the granular cells. They are rapidly, 
in less than a minute, swept to the intergranular lanes. On a slower time scale, say roughly 
30~minutes, the field lines  are further transported to the boundaries of our ``mesogranular'' 
network \citep{Stein:2006qy}. In our simulations different systems of field lines which 
have similar connectivity as they 
appear in the photosphere suffer reconnection amongst each other in the photosphere
and lower chromosphere. After half an hour or so the flux is established in the ``mesogranular''
flow and the different systems are separated into discrete intermittent positions at the 
boundaries of mesogranular scale cells, see figure~\ref{fig:3dr}. 

In the view seen from above (right panel, figure~\ref{fig:3dr}) the rotation of the originally 
{\em x}-oriented pre-existing ambient field in the corona is clear while comparing with figure~\ref{fig:3dri}. When the expanding  
portion of the flux tube reaches the corona, the direction of the field changes 
across the interface of the two flux systems (flux tube -- pre-existing, ambient).
The lower lying, deeper, lines (drawn in red) are more oriented in the same {\em y}
direction as the emerging flux.

\begin{figure*}
   \includegraphics[width=0.98\textwidth]{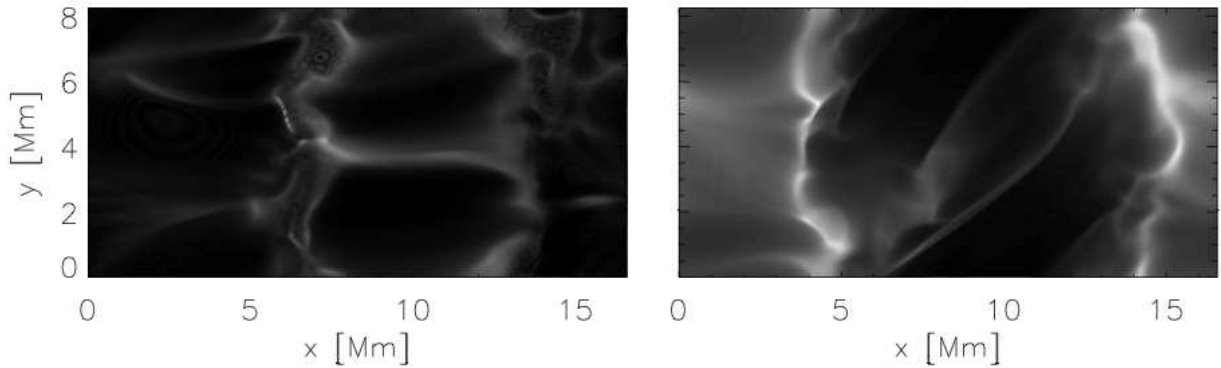}
  \caption{\label{fig:o6ca} Disc center synthetic O~{\sc vi} image at times $700$~s (left panel) and  $5280$~s (right panel).}
\end{figure*}

The low density structure (shaded in red) is oriented in the $xy$ (diagonal)
direction, evidence that the emerging flux and the ambient field have reconnected. 
This orientation is also evident in emission lines formed
in the transition region and corona. An example of this is shown in  
figure~\ref{fig:o6ca} where the simulated intensity of the O~{\sc vi}~$103.2$~nm emission line,
formed at roughly $3\,10^5$~K, is displayed at two different times, before and after the
flux expands into the corona. The O~{\sc vi} emission after flux emergence (right panel) 
clearly shows a large dark hole 
where the low density region is located. The high intensity regions of this line
outlines the footpoints of hotter loops which are located deeper in the atmosphere where the
densities are relatively large. In the A2 simulation we find that the low density region 
of the emerging flux found in the corona does not disappear; instead it increases in volume with
time because the low density region expands horizontally while the downflow
towards the loop foot points continues.

Late in the simulation, at time $4100$~s, small high density structures
are pulled into the chromosphere with the expanding flux, these structures are 
associated with the chromospheric portion
of the emerging flux tube (drawn in red) discussed above. Parts of these structures are
visible in figure~\ref{fig:3dr}: below the large low density region (shaded red), 
high density (green-yellow) is visible at greater heights than in the surrounding atmosphere.
After attaining a height of some $5$~Mm above the photosphere, the structures cease to rise, 
and while some of the plasma is trapped in magnetic dips and remains at great height, other 
plasma falls back towards the foot points (beginning roughly at $t=4400$~s). Dip like 
field configurations are produced naturally by the twist of the tube as noted by
\citet{Aulanier:1998cq,Lopez-Ariste:2006xq} and \citet{Magara:2007pt}. 
The structures grow horizontally thinner as the flow drains to the foot points 
with a velocity around $20$~km~s$^{-1}$ at $z\approx 4.1$~Mm. As they drain, the density 
contrast with the ambient chromosphere gradually fades and the structures are not 
discernible after $t=5000$~s.  

\begin{figure*}
   \includegraphics[bb=20 45 480 200,width=0.975\textwidth]{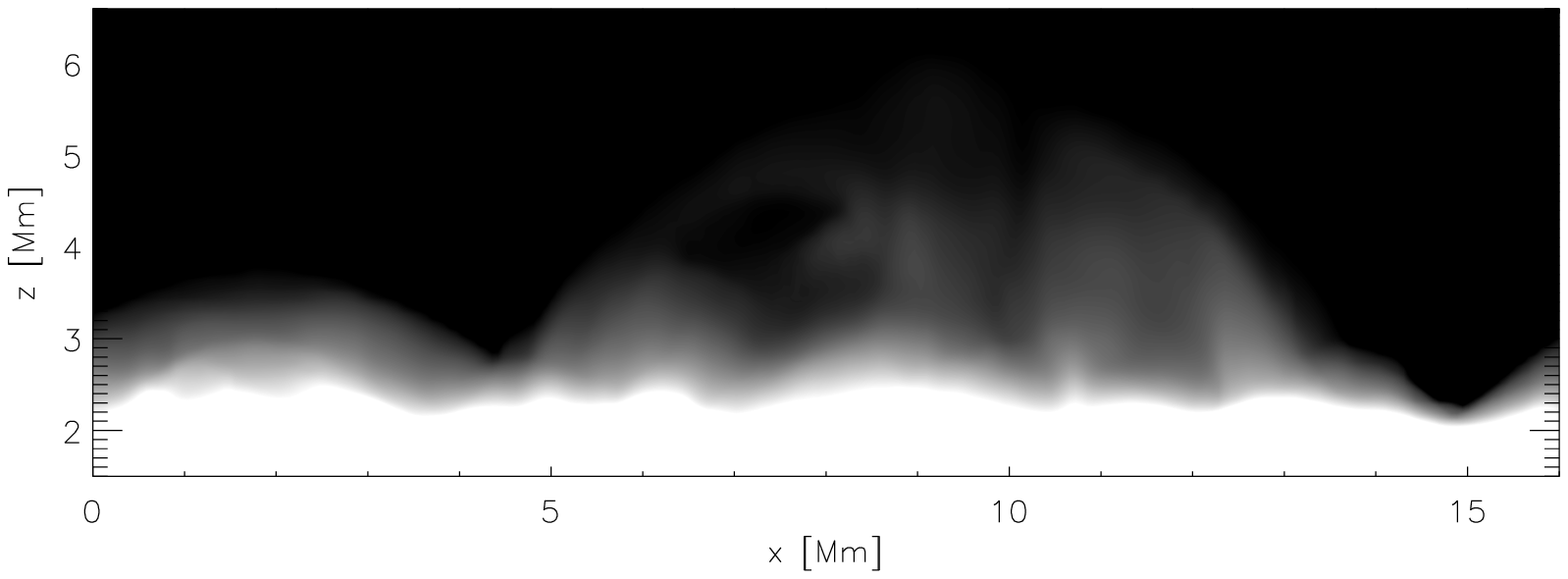}
      \includegraphics[bb=30 10 460 400,width=0.475\textwidth]{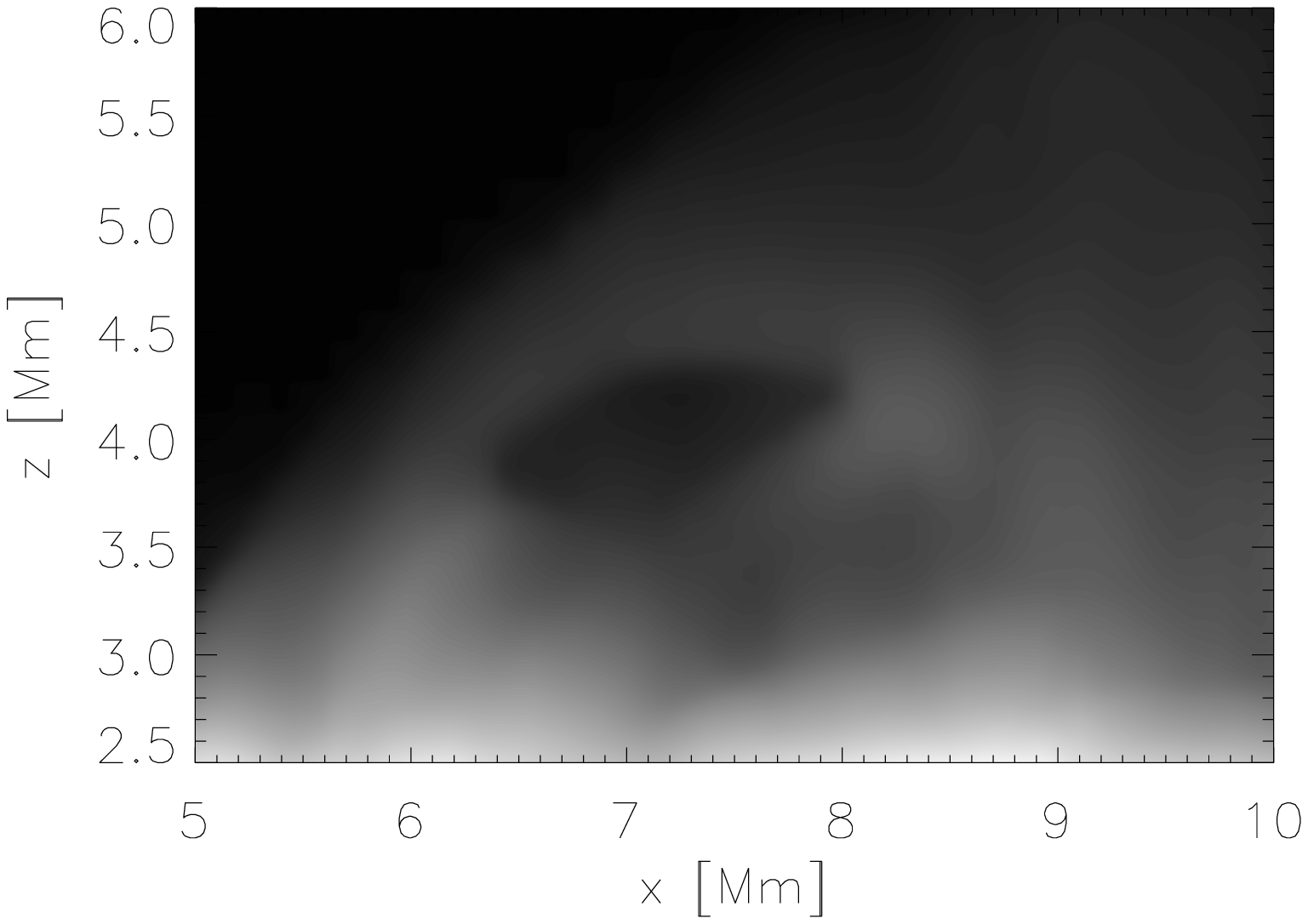}
      \includegraphics[bb=20 10 450 400,width=0.475\textwidth]{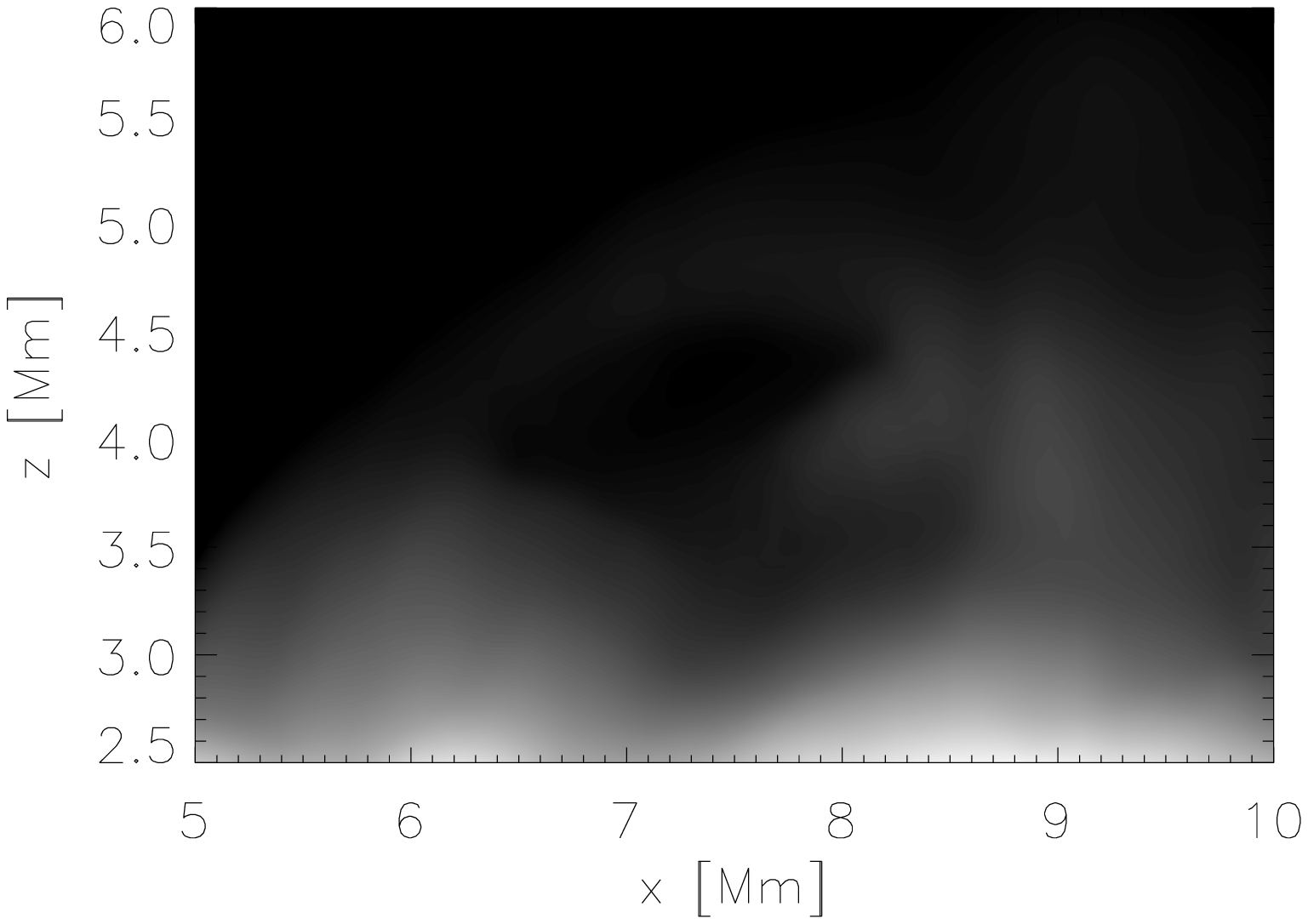}
  \caption{\label{fig:calim} Ca~{\sc ii} H-line synthetic image at the limb at time $4640$~s (top panel) and a close-up showing the 
evolution of a dark feature at time $4610$~s (bottom left) and at time $4640$~s (bottom right).}
\end{figure*}

Synthetic Ca~{\sc ii} H emission calculated at the limb (figure~\ref{fig:calim}) show 
additional dynamics occurring in these high density structures. Evident is a dark structure 
which is moving upwards while expanding. The darkening is due to a small bubble, or low
density inclusion, in the otherwise high density structure. The bubble is buoyant and 
moves upwards at roughly $10$~km/s. This low density bubble expands as mass 
drains from the upper layers of the high density structure surrounding it.

\subsection{Joule-Heating}
\label{sec:rec}

\begin{figure*}
  \includegraphics[width=0.98\textwidth]{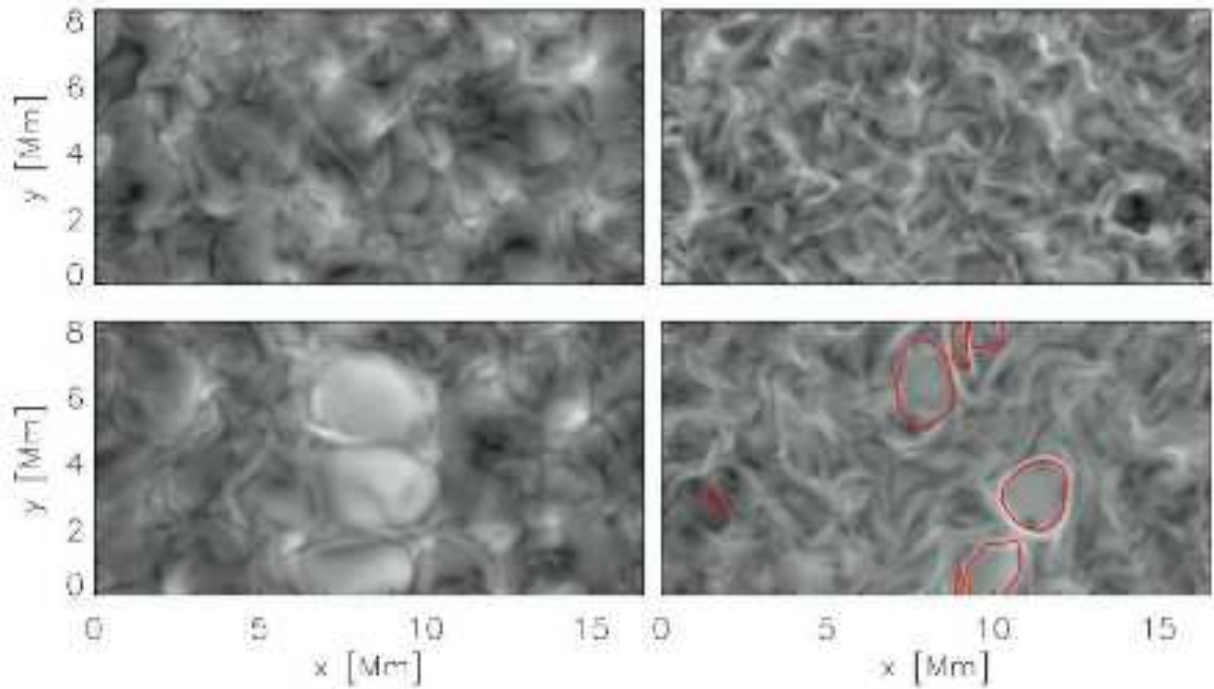}
  \caption{\label{fig:joulheat} Joule heating in a logarithmic scale shown at $z=458$~km (left panels) and at $z=1100$~km (right panels) 
    at times before the flux-tube reaches the photosphere at time $500$~s (top panels) and after the effects of the rising
    flux are evident at $2200$~s (bottom left panel) and $3390$~s (bottom right panel).
    Bright color corresponds to strong Joule heating. In order to show where the cold bubbles are, the bottom right panel includes a red contour where the temperature is 2300~K.}
\end{figure*}

As the emerging flux expands into the chromosphere and corona the Joule heating rate 
increases. The increase of the heating rate with height closely follows the rise of flux 
as described in figure~\ref{fig:bycor}. While crossing the upper photosphere the spatial 
structure of this heating follows the magnetic field concentrations and is similar
to  the granular cell pattern as shown in figure~\ref{fig:joulheat} (left panel).
The heating increases by more than a factor 3 in our simulations when the emerging flux crosses this 
height (458~km), but this is not enough to alter the temperature significantly.
The increase in heating rate depends critically on the amount of magnetic flux 
that reaches the photosphere. In simulation A2 twice as much Joule heating occurs 
at this height as compared to simulation A1: the time scale for Joule heating 
$\tau=e/(\eta j^2)$ is roughly $\tau=6000$~s for simulation A1 and $\tau=2500$~s for A2, 
during the time the major part of the Joule heating is due to the emerging flux. 
Thus, the greater the amount of magnetic flux that crosses the photosphere the greater 
the Joule heating. 

The structure of the Joule heating is different in the chromosphere: the most important 
location of Joule heating is at the boundaries of the large cold magnetized bubbles that 
rise into the chromosphere and were described in Paper~I. This is evident in the ellipsoidal
structures that fit the boundaries of the cold bubbles  (see the bottom right panel of 
figure~\ref{fig:joulheat}). This heating pattern appears at the same time as the cold bubbles 
in the chromosphere, {\it i.e.} at $t=3100$~s, some $1250$~s after the tube has crossed the 
photosphere in model A2. The Joule heating associated with the flux expansion is more 
than a factor $10$ greater than that given by the pre-existing ambient field. 
It is also more energetically important at this height than in the upper photosphere: 
the time scale for Joule heating is $\tau=550$~s for simulation A1 and is $\tau=300$~s for A2. 

\begin{figure}
  \includegraphics[width=0.48\textwidth]{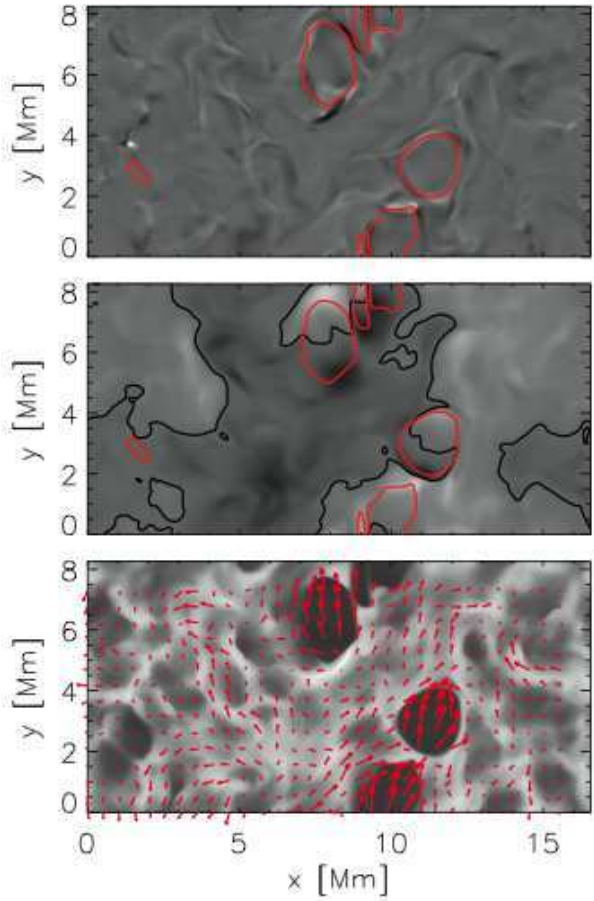}
  \caption{\label{fig:vrtz} $z$-omponent of the vorticity (top panel),
    $z$-component of the magnetic field (middle panel) and the
    temperature (bottom panel) at $z=1100$~km and time
    $3390$~s. The red contours show where the
    temperature is 2300~K (top and middle panel) outlining
    the cold bubbles, the black
    contour is $B_z=0$ (middle panel) and the red vectors show
    the horizontal field direction (bottom panel).}
\end{figure}

The bubbles start to appear some $1100$~s after the tube crosses the photosphere 
and slowly move through the chromosphere during the next $1500-2000$~s. 
We find that the bubble shape becomes progressively rounder
with greater magnetic field. Properties of the rising cold bubbles are shown in
figure~\ref{fig:vrtz}. The vertical field is positive in one half of the bubble and 
negative in the other half (middle panel). The horizontal magnetic field, on the other hand,
is unidirectional across the bubble (lower panel) and the horizontal component of the field 
is larger than the vertical component. Inside the bubbles the mean ratio between the 
horizontal and vertical magnetic field, at $z=1100$~km, is 25.7 while the range 
is $[4081.8,0.23]$. 
As discussed above, we find that Joule heating is concentrated near the boundaries of the 
bubbles. This turns out to be true also for the vorticity, which is large
in the vicinity of bubble boundaries (top panel). As time progresses the bubbles start 
to lose volume and become more elongated as the flux expands to greater heights and/or reconnects
with the ambient field. After some 4800~s of run A2, the magnetically dominated cold bubbles 
are fully dispersed and the atmospheric state returns to a shock dominated chromosphere
\citep[{\em e.g. } ][]{Carlsson:1992kl,Carlsson:1995ai,Carlsson:1997tg,Skartlien:2000lr,Wedemeyer:2004}. 

\begin{figure*}
\includegraphics[width=0.98\textwidth]{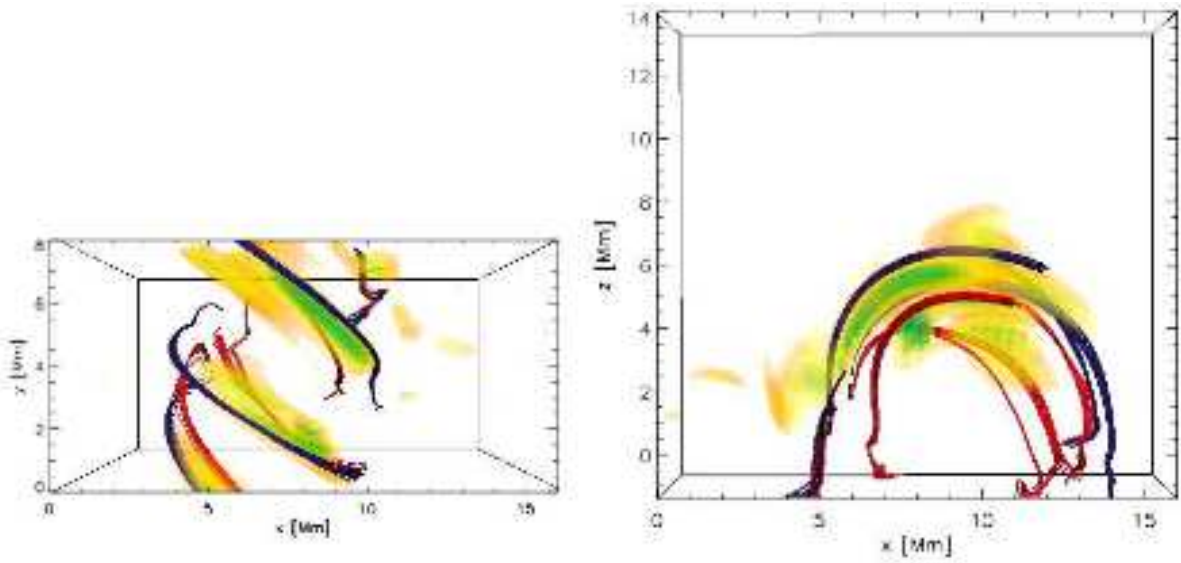}
\caption{\label{fig:3djou} Computational box from the convection zone to the corona is shown
  at time $4200$~s from above (left panel) and from the side (right panel). 
  Joule heating per particle is shown with yellow-blue color scale showing the banana structure. The magnetic field is shown above (blue) and below (red) two of these banana structures showing the discontinuity of the magnetic field.}
\end{figure*}

As we move further up into the upper chromosphere and lower corona, magnetic field discontinuity and the 
associated Joule heating becomes progressively more important. In figure~\ref{fig:3djou} we show
the Joule heating at $t=4200$~s in the upper parts of the 
simulated atmosphere. The structure of Joule heating resembles 
bananas. In the upper chromosphere and lower corona the time scale of 
Joule heating is quite short ($\tau=10$~s at height 2.5~Mm and at time $4480$~s of 
simulation A2). At these heights, the Joule heating is of the same order as the thermal
conduction and radiative losses.
The field lines (shown in red) below the Joule heating structures are oriented 
along the emerging flux tube whereas the field lines above (shown in blue) are oriented 
more along with the pre-existing ambient field. Most of the banana shaped Joule heating 
structures in the simulations show this kind of discontinuity in the field line orientation.

\begin{figure*}
\includegraphics[width=0.98\textwidth]{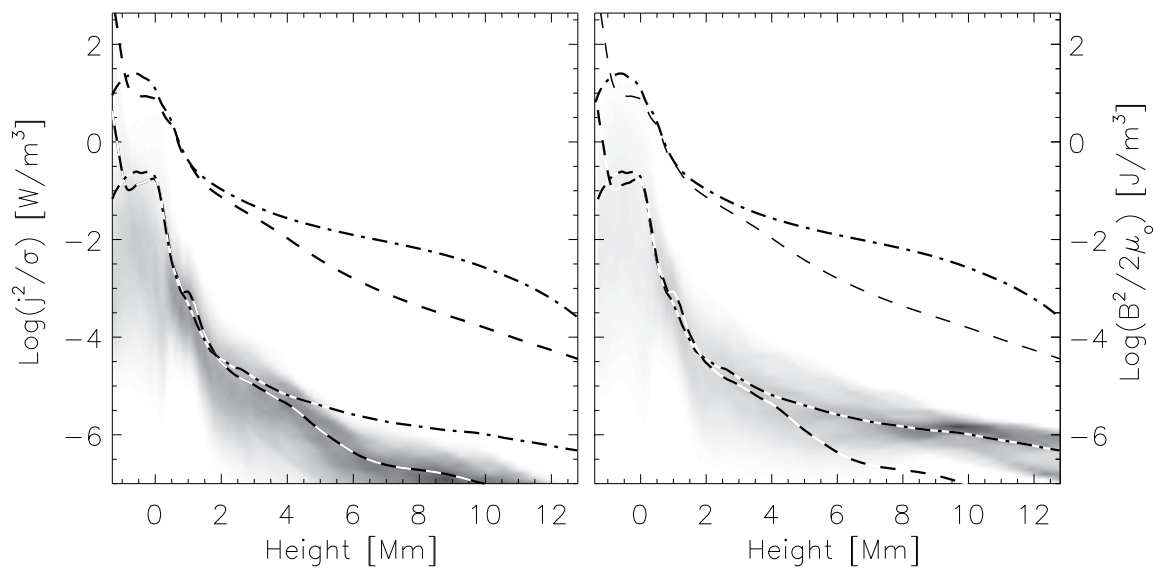}
\caption{\label{fig:pdfjou} Distribution (pdf) of Joule heating as function of height 
at time $t=1400$~s (left panel) and at time $t=5460$~s (right panel) of the simulation A2. 
Also shown are the average of the Joule heating at time $1400$~s (lower dashed line) and at time $5460$~s (lower dash-dotted line) and the average magnetic field energy ${B^2/2\mu_0}$ as a function of height at time $1400$~s (upper dashed line) and at time $5460$~s (upper dash-dotted line).}
\end{figure*}

It is not only the Joule heating rate in specific locations that increases with the 
introduction of emerging flux in the outer atmosphere, but also the general level of 
Joule heating throughout the upper chromosphere and corona. Figure~\ref{fig:pdfjou} shows 
the probability distribution function of Joule heating through the entire atmosphere as 
a function of height at two instances; $t=1400$~s and $t=5460$~s, the former before the 
emerging flux enters the outer atmosphere, the latter after the flux has expanded into the 
corona. At the earlier time the total magnetic energy (${B^2/2\mu_0}$) varies from 
0.5~J~m$^{-3}$ in the mid chromosphere at $z=1$~Mm to $1.25\,10^{-4}$~J~m$^{-3}$ high in the 
corona at $z=10$~Mm. Later, we find approximately the same average magnetic energy in the 
mid chromosphere, but it has increased greatly, a factor 20, in the corona to 
$2.5\,10^{-3}$~J~m$^{-3}$ at $t=5460$~s. The Joule heating rate increases in a similar manner. 
At any given height the heating rate can vary horizontally by $1-3$ 
orders of magnitude, while the average heating rate decreases exponentially with 
height. At the earliest time, $1400$~s, when only the pre-existing field contributes, 
the average Joule heating is found to be $8\,10^{-4}$~W~m$^{-3}$ at $z=1$~Mm, and less than 
$10^{-7}$~W~m$^{-3}$ in the corona at $z=10$~Mm. At the later time we find the same heating rate
at $1$~Mm, but the coronal heating rate at $10$~Mm has grown by an order of magnitude, to 
$8\,10^{-7}$~W~m$^{-3}$.

In order to isolate coronal heating, we integrate the Joule 
heating over the entire volume constrained by the box edges where the plasma 
temperature is greater than $10^5$~K (see top row of figure~\ref{fig:lineint}).
In simulation A2 (top middle panel), the total Joule heating
increases from $t=2000$~s to $t=3700$~s as the flux emerges into the upper chromosphere 
and corona and interacts with the pre-existing ambient field. This heating increase is mostly 
confined to the extensions of the rising cool bubbles discussed earlier.
From $t=3600$~s to $t=4200$~s the Joule heating in `banana-like' 
structures appears, and the total Joule heating increases and shows greater variations with time.
At $t=4700$~s there is a large jump in the total Joule heating in simulation A2. 
This large jump in the total Joule heating is related with the high temperatures close to 
the transition 
region ($>6\,10^5$~K shown in orange in the left panel of figure~\ref{fig:3dtj}).
Such blobs of hot plasma at the loop foot points near the transition 
region appear when the emerging flux comes in contact with the pre-existing ambient 
magnetic field, producing high Joule heating rate in a small region.
The blob of hot plasma is first 
heated in the transition region but the high temperature rapidly (in some $20$~s) 
propagates into the corona by conduction. 
This process is repeated several times in the same vicinity roughly every 3~minutes.
These oscillations are produced by the displacement of the transition region 
caused by the flux expansion which perturbs the atmosphere. This 
oscillatory movement stretches the flux tube lines with the 
ambient field lines, first in one specific region at one of the footpoint bands of the transition 
region and later at the opposite footpoint band. During
these events the maximum temperature in the corona varies from $5.9\,10^5$ to $6.8\,10^5$~K. 
The first event is the strongest, after the strength of the Joule 
heating events decrease. The time scale of the Joule heating is in the first instance 
some $\tau=4.5$~s.

\subsection{Coronal energy, temperature and Intensities}
\label{sec:enerline}

\begin{figure*}
  \includegraphics[width=13cm]{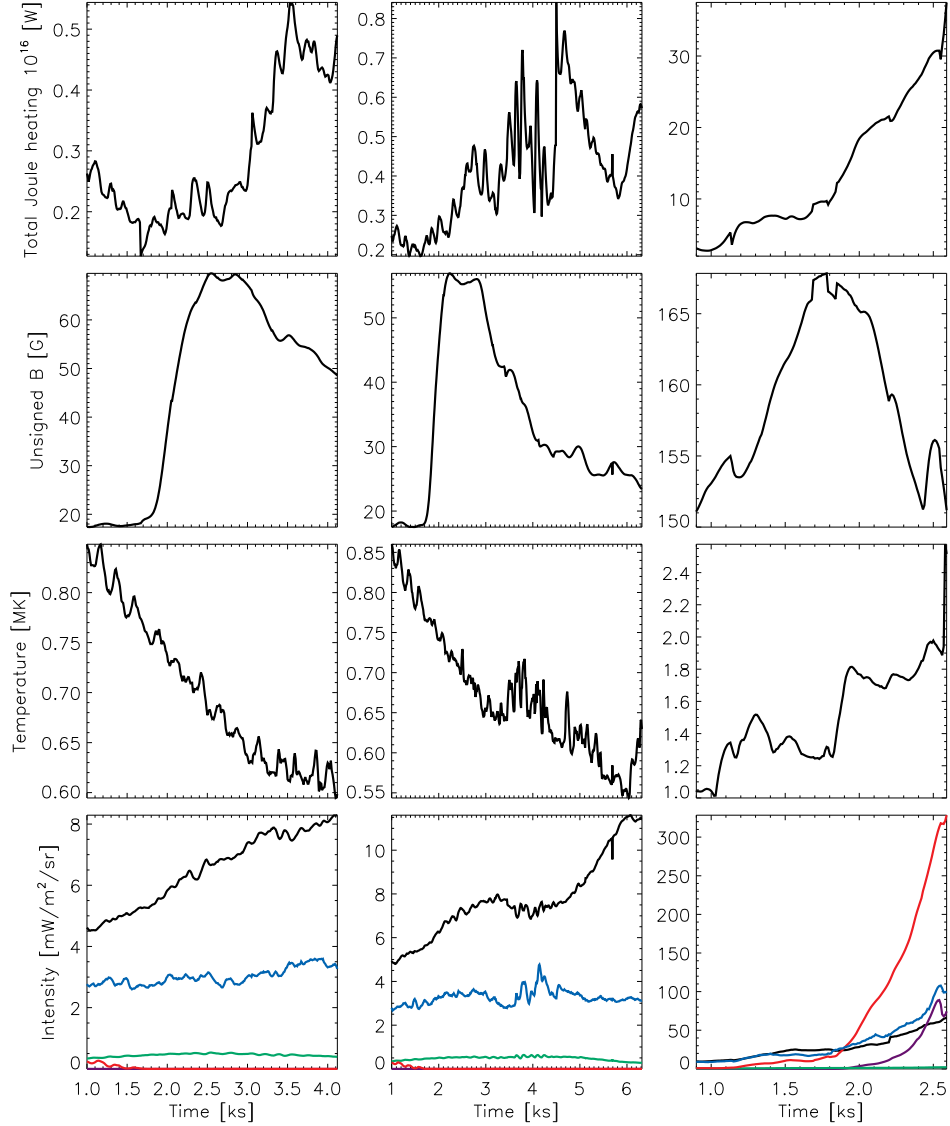}
  \caption{\label{fig:lineint} Integral of the Joule heating over the corona, {\it i.e.} the volume where the temperature is above $10^5$~K (top panels),  
unsigned magnetic field strength in the photosphere (second row of panels), maximum temperature (third row of panels) 
and mean emission of the He~{\sc ii} (blue line), O~{\sc vi} (black line), Ne~{\sc viii} 
(green line), Fe~{\sc xii} (red line) and Fe~{\sc xv} (violet line). Evolution shown for 
the simulation A1 (left column), A2 (middle column) and B1 (right column) as a function of time.}
\end{figure*}

\begin{figure*}
  \includegraphics[width=0.98\textwidth]{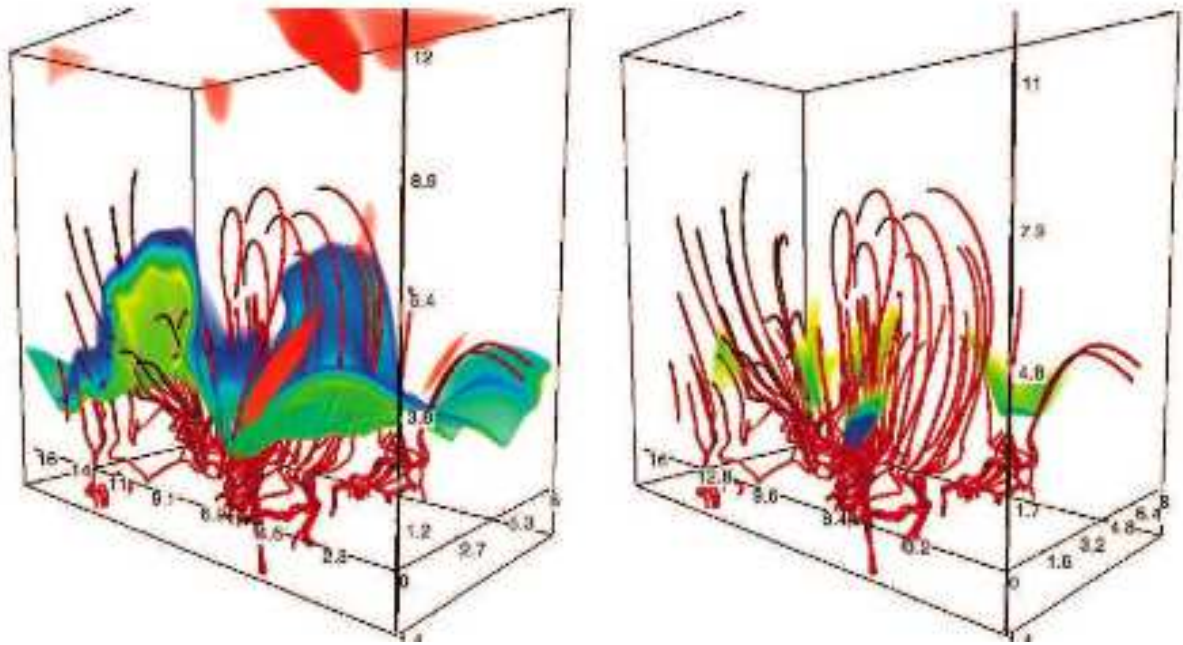}
  \caption{\label{fig:3dtj} The computational box from the convection zone to the corona at 
time $4720$~s. The temperature in the transition region is shown with a yellow-blue color scale
(left panel) with the highest temperatures shown in orange (up to $6\,10^5$~K). 
Joule heating is shown with a yellow-blue color scale (right panel). Both images include  
magnetic field lines shown in red.}
\end{figure*}

To further investigate the variation of the Joule heating in time and
its possible effects on coronal energetics and diagnostics we return
to figure~\ref{fig:lineint} where we show the total Joule heating
rate, along with the variation of the average photospheric magnetic
field, the maximal coronal temperature, and some specific synthesized EUV
resonance lines observable with spectrographs such as Hinode/EIS and
SOHO/Sumer, as functions of time.

Let us first concentrate on simulation A1 (left column). The 
maximum temperature (third panel from the top) falls in time, indicating that the entire corona is cooling. The maximum
temperature falls from some $8.5\,10^5$~K to $5.5\,10^5$~K from $t=1000$~s to $t=5700$~s. This is due
to the imbalance between the heating and cooling processes; the Poynting flux being dissipated in
the corona is not sufficient to balance thermal conduction and radiative losses. The paucity of
coronal heating is primarily due to the weak pre-existing ambient field in the model. 
The average unsigned magnetic field strength in the photosphere is initially 
roughly $15$~G in runs A1 and A2. It stays almost constant until $t=1500$~s and then 
increases quickly up to $50$~G (A2) or $55$~G (A1); then stays constant for another 800~s. 
Afterwards, the field strength in the photosphere decreases to $25$~G for simulation 
A2 (at $5500$~s) and to $40$~G for simulation A1 (at $3500$~s). In the simulation 
B1 the field strength is initially $150$~G but increases to $162$~G (at $1500$~s), 
thereafter going down to $152$~G (at $2500$~s).
We also find falling temperatures in the A2 simulation. However, note that in both 
simulations the drop in maximum temperature slows as the Joule heating increases 
at around $t=3500$~s. Note that the flux emerges into the corona
already at $t=2000$~s, and needs some time to interact with the ambient field before 
any increase of the temperature is observed.
In contrast to the runs A1 and A2 we find that in run B1 the maximum temperature does 
not decrease with time, but rather increases, from $1$~MK at $t=500$~s to nearly 2~MK 
at $t=2500$~s. This is presumably due to the much stronger pre-existing ambient field in 
this simulation (8 times larger than in the A1 and A2 simulations). In addition it is 
worth mentioning that the effects of flux emergence on the maximum coronal temperature 
occurs more rapidly in this run, almost immediately after the emerging flux enters into 
the corona.

In the bottom row of figure~\ref{fig:lineint} we show the synthesized total intensity in 
various EUV spectral lines typical of different temperature regimes in the corona as a 
function of time. The synthesized intensity of these lines is computed assuming the 
optically thin approximation and may be written as

\begin{equation}
I={h\nu\over 4\pi}\int_sn_{\rm e}n_{\rm H}g(T_{\rm e})ds
\end{equation}                  

\noindent where the integration is carried out along the line of sight, or in this case vertically 
through the box. 

The He~{\sc ii} $25.6$~nm line is formed at some $8\,10^4$~K and is a 
typical lower transition region line, its total intensity is shown in blue in the bottom row of figure~\ref{fig:lineint}. We find that 
the intensity is roughly constant in the A1 and A2 simulations, though in the A1 run there 
is a slight increase in the intensity during the run and strong oscillations in the A2 run when the emerging flux perturbs the transition region. In the B1 run the He~{\sc ii} line intensity 
is initially of the same order, less than 10~mW~m$^{-2}$sr$^{-1}$, but 
at the end of the run the intensity has risen quite substantially to almost 100~mW~m$^{-2}$sr$^{-1}$. 
The O~{\sc vi} 103.2~nm line (drawn in black) formed at some $3\,10^5$~K rises from 
$5$~mW~m$^{-2}$sr$^{-1}$ and essentially doubles during the A1 and A2 simulations. 
In the B1 run the O~{\sc vi} line rises from $5$ to 
$60$~mW~m$^{-2}$sr$^{-1}$ during the time shown. The Ne~{\sc viii} $77$~nm line (drawn in green) 
is formed at $6.3\,10^5$~K, but remains weak, of order $1$~mW~m$^{-2}$sr$^{-1}$, in all three simulations.

Finally, we show two coronal lines, Fe~{\sc xii} 19.5~nm formed at 1.25~MK and 
Fe~{\sc xv} 284~nm which is formed at 2~MK. 
These lines are negligible in the A1 and A2 runs but increase sharply with rising coronal temperatures
in the warmer B2 simulation, where the radiative losses from Fe~{\sc xii} rise from $4$ to $20$~mW~m$^{-2}$sr$^{-1}$ at $1600$~s, afterwards increase rapidly to greater than $300$~mW~m$^{-2}$sr$^{-1}$ at $2300$~s, while the hot Fe~{\sc xv} line emits some $100$~mW~m$^{-2}$sr$^{-1}$ at the 
end of the simulation. 

Note that the Fe~{\sc xv} line follows the maximum temperature variation
fairly accurately towards the end of the simulation with strong oscillations from $30$ to $90$~mW~m$^{-2}$sr$^{-1}$ after $2200$~s. This is generally not the case, presumably 
because the total line intensity depends on both the density of the plasma emitting at the 
temperature of line formation and on the volume of emitting gas along the line of sight, which
depends on the temperature gradient and/or topology of the corona. The relatively slow response 
of the coronal lines may reflect the significant timescales of heating or cooling significant
amounts of material to or from coronal temperatures.

\section{Discussion and Conclusions}
\label{sec:conclusions}

Magnetic flux tube emergence is obviously important for the evolution of the
magnetic field in the outer solar layers, but it is also of major significance
in determining the dynamics and energetics of the chromospheric and coronal
plasma. 

We find that, after initially stalling in the photosphere,
the emerging flux tube fragments, allowing discrete magnetized bubbles to 
expand into the chromosphere. It is almost impossible to follow the 
axis of the flux tube or its apex since the tube is shredded by plasma dynamics 
and mixes with the extant ambient magnetic field. This makes it difficult 
to study the velocity of the flux tube or its exact position. Nevertheless,
what is clear is that the greater part of the flux tube is stuck in the 
upper photosphere; {\it i.e.} we do not observe any flux rope emergence 
in the regions above the photosphere. However, as mentioned, the upper 
parts of the magnetic flux tube expand into the corona. This expansion pushes 
the transition region upwards to at least $6$~Mm above the photosphere 
and also ensures that the regions of the upper chromosphere which contain the 
strongest field maintain a low density.

As the flux from the emerging tube crosses the chromosphere and corona, 
it interacts and eventually merges with the pre-existing ambient field. 
The resulting magnetic field configuration is a combination of the extant 
and newly emerging fields such that the direction of the resulting field 
lies between these two systems.  The field from the emerging flux tube that 
rises into the corona also interacts with the pre-existing open field lines. 
These open field lines close and are seemingly pulled downward into the 
corona as at the same time the field lines from the expanding portion of 
the emerging field are moving upward. We find a low plasma density structure 
situated between these two systems of field lines. The low density structure 
is robust and lasts for more than thirty minutes. It is produced by a 
downflow from the lower corona to the photosphere
following the field lines. The total pressure in the low density region is lower
than the total pressure in the field line systems above and below resulting in
them getting closer to each other. The pressure imbalance is partly offset by
magnetic tension forces which slows the process. 

Just below the low density structure we find that a high density structure 
forms in the chromosphere. This structure rises with the expanding field lines 
and is maintained at above hydrostatic heights by a dip-like structure of 
the magnetic field that is produced naturally by the twist in the expanding 
field. The structure thins with time as portions of its mass drain to the 
foot points of the supporting magnetic field lines. The structure described
is in some ways suggestive of solar quiescent filaments even though 
filaments are much larger scale structures that are, as yet, impossible to 
simulate with the type of simulation described here. 

Associated with the high density structure we also find low density 
inclusions that display interesting dynamics shown in the synthetic 
Ca~{\sc ii} images at the limb that resembles dynamics observed in quiescent 
prominences
with the Hinode spacecraft  \citep{Berger:2008qf}. The low density structure 
moves upwards at buoyant velocity and expands as material drains out of the
regions above it. Even so, we are hard pressed from finding counterparts to
all of the dynamic phenomena observed at the solar chromospheric
limb with Hinode or the Swedish 1-meter Solar Telescope.

The emerging field lines that remain in the lower chromosphere become
more tangled than those that expand all the way into the corona where 
the field has more or less uniform direction. The foot points of
the various field lines systems are connected intermittently in the 
network situated in the boundaries of the ``mesogranular'' scales discussed 
in Section~\ref{sec:structure}. The field from different heights seems 
to concentrate in different intermittent ``mesogranular'' regions which all
evolve on timescales typical of the convective layers below the photosphere.

One of our goals is to study the Joule heating produced by interaction between the emerging flux
and the pre-existing ambient magnetic field. In these simulations we have run two
cases with weak pre-existing magnetic field and fairly strong emerging field 
and one case with a stronger pre-existing field and a weak emerging field.
In all cases the Joule heating increases throughout the atmosphere as the 
emerging field expands, but it seems that the additional Joule heating 
is more vigorous when the pre-existing field is stronger.

The Joule heating caused by emerging flux is localized and surrounds the 
cool magnetized bubbles that rise into the chromosphere, and at later times 
form elongated structures. 
Joule heating is energetically more important in the chromosphere than in 
the photosphere and increases in importance at even greater heights to become 
a dominant term in the energetics in the corona. The structure of the 
Joule heating in the upper-chromosphere and corona takes on banana shaped forms
that follow the magnetic field lines. 

Joule heating has both a quasi-stationary component as well as occurring in 
episodic events of larger amplitude. The latter is evident in the 
corona as high temperature blobs are produced, as a rule near the transition 
region interface. Thus, near the foot points of a coronal loop, 
high temperature gas is formed in discrete bursts. The injected heat moves 
into the higher layers of the corona as a result of 
thermal conduction, on short timescales. Such energetic 
episodic events are often repeated several times, in the same place, but 
they becomes weaker with each repetition.

Flux emergence changes the total emission of various transition region and 
coronal lines. In general the emission of O~{\sc vi} and Ne~{\sc viii} 
increases as emerging flux expands into the corona. However, in the runs 
with a weak pre-existing magnetic field the temperature remains too low
to give rise to emission in coronal lines such as Fe~{\sc xii} or Fe~{\sc xv}.
In the strong field case these lines increase their intensity very rapidly, 
at the end of the run the Fe~{\sc xii} is the strongest line of those 
modeled. In no case do we find a one-to-one correspondence between the
variations of the temperatures and the line intensity of the coronal lines.
When the corona is heated more matter is brought to coronal temperatures. In addition,
the density rises which occurs on a different, generally longer, 
timescale than the heating timescale which governs the evolution of the temperature. 
The resulting line emission reflects a combination of these time scales.

\section{Acknowledgments}

This research has been supported by 
a Marie Curie Early Stage
Research Training Fellowship of the European Community's Sixth Framework
Programme under contract number MEST-CT-2005-020395: The USO-SP International School for Solar Physics. 
Financial support by the European Commission through the SOLAIRE Network (MTRN-CT-2006-035484) and by the Spanish Ministry of Research and Innovation through project AYA2007-66502 is gratefully acknowledged. 
This research was supported
by the Research Council of Norway through grant 170935/V30 and through grants of computing time from the Programme for Supercomputing. 
To analyze the data we have used IDL and Vapor 
(http://www.vapor.ucar.edu).
Sven Wedemeyer-B\"ohm, Jorrit Leenaarts and Chung Ming Mark Cheung are thanked for valuable discussions and comments.

\begin{deluxetable}{ccccccccc}
\tablecaption{\label{tab:runs} Summary of parameters for the three simulations}
\tablehead{
\colhead{Name} & \colhead{Twist $\lambda$}  & \colhead{B$_0$ [G]} & \colhead{Size [$Mm^3$]}  & \colhead{Time [s]} & \colhead{Radius [Mm]} & \colhead{Ambient field [G]}}
\startdata
A1 & 0.3 & 4484 & $16\times 8\times 16$ &  4200 &  0.5 & 16 \\ \\
A2 & 0.6 & 4484 & $16\times 8\times 16$ &  6300 &  0.5 & 16 \\ \\
B1 & 0 & 1121 & $16\times 8\times 16$ &  2600 &  1.5 & 155 \\ \\
\enddata
\end{deluxetable}

\end{document}